Microscopic Mechanism of Carbon Annihilation upon SiC Oxidation due to Phosphorous Treatment: Density-Functional Calculations Combined with Ion Mass Spectrometry

Takuma Kobayashi,[1)] Yu-ichiro Matsushita,[2)] Takafumi Okuda,[1)] Tsunenobu Kimoto,[1)] and Atsushi Oshiyama[2)]

[1]Department of Electronic Science and Engineering, Kyoto University, Katsura, Nishikyo, Kyoto 615-8510, Japan

[2]Department of Applied Physics, The University of Tokyo, Hongo, Bunkyo-ku, Tokyo 113-8656, Japan

We report first-principles static and dynamic calculations that clarify the microscopic mechanism of carbon annihilation due to phosphorous treatment upon oxidation of silicon carbide (SiC). We identify the most stable form of the phosphorus (P) in the oxide as the four-fold coordinated with the dangling PO unit and find that the unit attracts carbon ejected from the interface, thus operating as a carbon absorber. This finding provides a microscopic reasoning for the first time for the promotion of the oxidation reaction on one hand and the annihilation of the C-related defects at the interface on the other. Secondary ion mass spectrometry measurements are also performed and the obtained carbon profile corroborates the theoretical finding above.

The interface of a condensed matter and its oxidized film usually formed by the oxidation is ubiquitous in nature and provides an important stage in both science and technology. In semiconductor devices, for instance, the interface becomes a channel for the electric current and the band offset at the interface allows the gate controllability of the transistor action [1]. Hence the identification of the atomic structure and the clarification of its electronic functionality of the interface are challenging and demanded in nanoscience and technology.

An example which has been insufficiently investigated in spite of its importance is the interface of silicon carbide (SiC) and silicon dioxide ($SiO_2$). SiC is a candidate material for the sustainable power electronics in future due to its superior properties such as wide bandgap, high critical electric field and low intrinsic carrier concentration [2,3]. The additional but important advantage of SiC is that $SiO_2$ films formed by the thermal oxidation can be used as gate insulating films in metal-oxide-semiconductor field effect transistors (MOSFETs) [3], ensuring the connectivity with the current Si technology.

However, were carbon atoms not annihilated during the oxidation, the $SiO_2$ films would not be realized. The fact is that, although SiC MOSFETs are fabricated, the mobility of the devices is lower than that of SiC bulk by two orders of magnitude [3]. This is certainly due to the interface state density, $D_{it}$, which is typically about $10^{13}$ - $10^{14}$ $cm^{-2}eV^{-1}$ [3,4,5] near the conduction-band edge ($E_C$) of SiC which is higher by more than three orders of magnitude than that in typical $SiO_2$/Si systems ($D_{it} \sim 10^{10}$ cm$^{-2}$eV$^{-1}$) [6]. The origin of those interface states is believed to be carbon related defects [3,5,7,8].

Introduction of foreign atoms such as phosphorus (P) [9], boron (B) [10], sodium (Na) [11], and barium (Ba) [12] into SiO$_2$/SiC systems reduces the $D_{it}$ (passivation of the interface levels), thus leading to the increase of the channel mobility. It is experimentally known that the foreign atoms mentioned above are distributed in the SiO$_2$, and that the oxidation reaction is promoted during the passivation process. These observations along with the chemical diversity of the introduced passivating foreign atoms are mysterious, and indicate that passivation through the direct attachment to the interface defect, as in the hydrogen attachment to the Si dangling bond at the SiO$_2$/Si interface (passivation of the P$_b$ center) [6], is unlikely.

In this Letter, we report static and dynamic calculations based on the density-functional theory (DFT) [13] that provide a microscopic mechanism of the passivation of C-related defects at the SiO$_2$/SiC interface for the first time. We find that the P atom resides in its peculiar form in SiO$_2$ and operates as a C absorber, thus annihilating the C-related defects at the interface. We also report our secondary ion mass spectrometry (SIMS) measurements that corroborate the theoretical finding.

All calculations are performed in the DFT using our real-space scheme (RSDFT code) [14,15] in which Kohn-Sham equations are discretized on three-dimensional spatial grids and solved by the real-space finite-difference method [16,17]. The generalized gradient approximation by Perdew,

Burke and Ernzerhof (PBE) [18] is adopted for the exchange-correlation energy. Nuclei and core electrons are simulated by norm-conserving pseudopotentials [19]. We find that the grid spacing in the real space of ~ 0.16 Å, corresponding to the cutoff energy of ~ 82 Ry in the plane-wave basis set, suffices to ensure the accuracy within ~ 0.15 meV/atom in the total-energy difference. We have used sufficiently large supercell model so that the Γ point sampling is enough for the Brillouin-zone integration. In the structural optimization, the force acting on each atom is minimized less than 26 meV/Å. In the dynamic calculations at finite temperature, we adopt Car-Parrinello molecular dynamics (CPMD) scheme [20] implemented in RSDFT code and perform constant volume and temperature (NVT) simulations.

The phase diagram experimentally determined for the interaction of oxygen with the SiC surface [21] indicates that most of the carbon atoms are ejected from the interface as forms of CO units during the oxidation of SiC and diffuse to the $SiO_2$ side. Recent DFT calculations [22] have clarified that C atoms which form carbon clusters are also ejected from the interface as CO. Hence, to examine the mechanism of the passivation, we consider the fate of the CO unit in P-doped $SiO_2$.

We now set out with exploring the structure of P-doped $SiO_2$. Upon the annealing with $POCl_3$ usually with $O_2$ and $N_2$ [9], $SiO_2$ is known to change its structure to phosphosilicate glass (PSG) in which P is incorporated in $SiO_2$ with the concentration of typically $10^{21}$ cm$^{-3}$. In the annealing with $POCl_3$, it is speculated that P becomes the form of $P_2O_5$ [23]. To identify the microscopic structure of PSG, we have then performed a CPMD simulation for a system consisting of $\alpha$-quartz $SiO_2$ (20 molecular units) and 2 $P_2O_5$ units. In the simulation, the target system was first annealed at about 4000 K. By this annealing, sharp peaks disappeared from the radial distribution function, indicating that the structure completely turned into liquid-like. Subsequently, the structure was cooled down to 1000 K in the rate of about $-50$ K/picoseconds, presumably becoming a typical amorphous form. As a result of the first-principle CPMD simulation, we have found common structural characteristics in which the P atom is four-fold coordinated with O atoms and one of the four PO bonds is dangling toward the interstitial region as shown in Fig.1 (an $-O_3PO$ configuration hereafter).

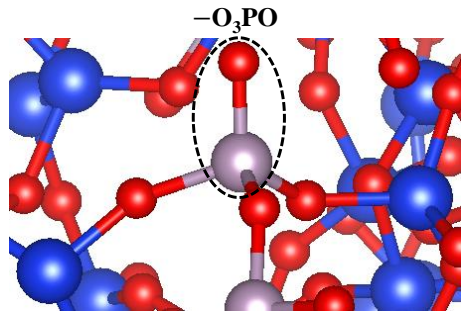

Fig.1: Typical $-O_3PO$ configuration obtained by Car-Parrinello molecular dynamics (CPMD) simulation for phosphosilicate glass (PSG). The blue, red and silver balls denote Si, O and P atoms, respectively.

We are then in a position to clarify the most energetically favorable configuration of the CO unit which is ejected from the interface toward P-doped $SiO_2$, i.e., PSG. Since we have identified the most stable form of the P atom in $SiO_2$ as the $-O_3PO$ configuration (Fig.1), we prepare a 73-atom simulation cell in which an $-O_3PO$ unit exists in 23 $SiO_2$ molecular units. When the $-O_3PO$ unit is introduced in SiO$_2$, there comes up a Si dangling bond which is terminated by an H atom. Then we introduce a CO molecule in the interstitial region as in Fig.2(a). We have optimized the geometry of this configuration and found that the CO molecule stays in the interstitial region.

We have then moved the CO molecule in the vicinity of the –O$_3$PO configuration and found that the CO is eventually trapped by the –O$_3$PO and forms a configuration in which the dangling PO unit is replaced by a PCO$_2$ unit (Fig.2(b)), denoted as –O$_3$PCO$_2$ hereafter. This reaction is found to be exothermic with the energy gain of 1.3 eV. This theoretical finding indicates that the P atom doped by the POCl$_3$ annealing in SiO$_2$ operates in its peculiar structural form, –O$_3$PO, as a carbon absorber. This may promote the carbon removal from the SiO$_2$/SiC interface, thus decreasing the interface state density $D_{it}$ on one hand and promoting the oxidation reaction on the other.

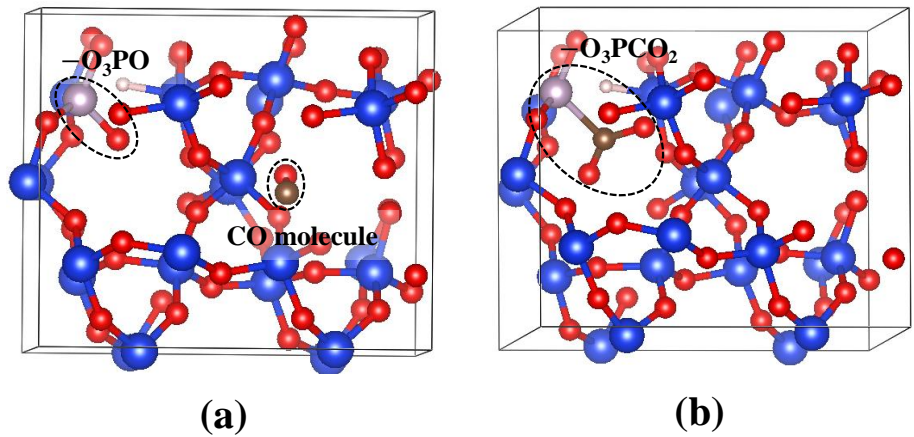

**(a)** **(b)**

Fig.2: Structures of CO-injected phosphosilicate glass (PSG) obtained by our structural optimization in DFT. (a) The structure with the CO molecule staying in the interstitial and (b) the structure after the formation of the –O$_3$PCO$_2$ configuration. The blue, red, silver, brown and white balls depict Si, O, P, C, and H atoms, respectively.

A remaining important issue is the stability of the $-O_3PCO_2$ configuration in $SiO_2$ in realistic situations of the SiC oxidation with the $POCl_3$ annealing. The system we consider is a polyatomic set under a certain pressure $P$ at finite temperature $T$ where each atomic species diffuse into or out of the system under a certain chemical potential $\mu_i$ of the $i$-th atomic species. Then the free energy which should be compared is the Gibbs free energy, $G = E + PV - TS - \sum_i \mu_i N_i$, where $E$, $S$, $V$ and $N_i$ are the total energy, the entropy, the volume and the number of the $i$-th atom of the target system. First, the $-O_3PCO_2$ configuration may be transformed to the three-fold coordinated $-O_3P$ configuration plus an interstitial $CO_2$ ($-O_3P + CO_2$). Second, there is a possibility that the CO reacts with $O_2$ in $SiO_2$, becoming a form of $CO_2$. Then $CO_2$ may approach the configuration, $-O_3PO$, which is the most stable in PSG (i.e., $-O_3PO + CO_2$). This configuration may be also realized by further oxidation of the $-O_3PCO_2$ configuration. We have then considered the two competing configurations, the $-O_3P + CO_2$ and the $-O_3PO + CO_2$, and fully optimized the structures as shown in Figs. 3(a) and (b).

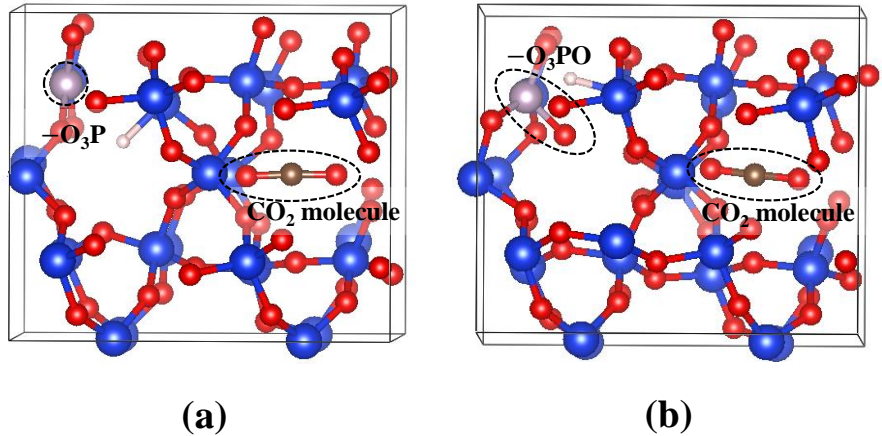

(a)          (b)

Fig.3: Structures of phosphosilicate glass (PSG) containing $CO_2$ molecule in the interstitial obtained by our structural optimization. (a) The structure with the $-O_3P + CO_2$ and (b) the structure with the $-O_3PO + CO_2$). The color code is the same as in Fig. 2.

For the comparison among the three configurations, the $-O_3PCO_2$, the $-O_3P + CO_2$ and the $-O_3PO + CO_2$, we need the chemical potential of oxygen. In the oxygen rich condition, we reasonably assume that the oxygen chemical potential is equal to that in an $O_2$ molecule: $\mu_O = \mu_O(O_2)$. We have calculated the total energy of an $O_2$ molecule and set $\mu_O(O_2) = E(O_2)/2$ at zero temperature. At finite temperature, we have obtained $\mu_O(O_2)$ from the thermochemical table [24]. In the oxygen poor condition, $\mu_O$ is considered to be the chemical potential in $SiO_2$, $\mu_O(SiO_2)$. From the calculated total energies of $SiO_2$ ($\alpha$-quartz) and Si crystal, we have evaluated $\mu_O(SiO_2)$ at zero temperature by $\mu_O(SiO_2) = (E(SiO_2) - \mu_{Si})/2$ with assuming $\mu_{Si} = \mu_{Si}(Si\ crystal)$. At finite temperature, the pertinent corrections are obtained again from the thermochemical table [24].

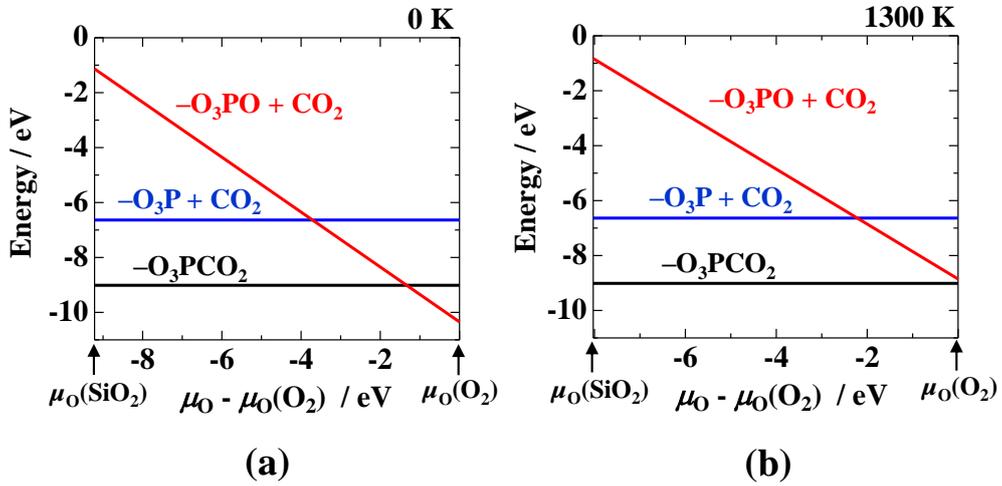

Fig.4: Comparison of the Gibbs free energies among three competing structures in phosphosilicate glass (PSG) related to the newly found $-O_3PCO_2$ configuration as a function of the oxygen chemical potential at (a) 0 K and at (b) 1300 K. The three structures, $-O_3PCO_2$, $-O_3P + CO_2$, $-O_3PO + CO_2$, are shown in Fig.2(b), Fig.3(a) and Fig.3(b), respectively. In (b), the standard-state pressure ($p^0 = 0.1$ MPa) is used for the oxygen pressure [25].

Figures 4(a) and (b) are the obtained free energy for the three configurations as a function of the oxygen chemical potential at 0 K and 1300 K, respectively. The temperature of 1300 K is about the typical temperature for the POCl$_3$ annealing [9]. We have found that the –O$_3$PCO$_2$ configuration (Fig.2(b)) is always lower in energy than the –O$_3$P + CO$_2$ configuration (Fig.3(a)) irrespective of the temperature. This means that the generated PC bond in the –O$_3$PCO$_2$ is unbreakable by the thermal annealing alone. We have also found that the –O$_3$PCO$_2$ is lower in energy than the –O$_3$PO + CO$_2$ (Fig.3(b)) for all the allowed value of $\mu_O$ at 1300 K. Hence it is highly likely that the ejected CO molecules from the interface are captured by the –O$_3$PO local configuration in PSG rather than react with O$_2$ to become CO$_2$ during the POCl$_3$ annealing. When we compare Figs. 4(a) and (b), the structure –O$_3$PCO$_2$ becomes more stable by increasing the temperature. Thus, if we perform the POCl$_3$ annealing at higher temperature than 1300 K, the amount of the remaining –O$_3$PCO$_2$ inside the PSG is predicted to increase.

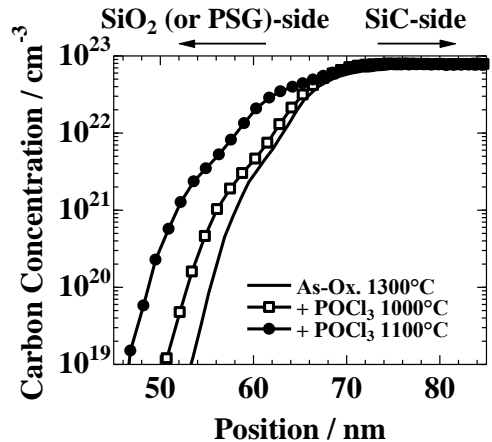

Fig.5: Evaluated profile of C concentration near the interface of SiC and the SiO$_2$ or phosphosilicate glass (PSG) films determined by the SIMS measurement. The abscissa, i.e. the position, for the three data are shifted so that the position where C concentration saturates (~ 10$^{23}$ cm$^{-3}$) in the SiC region becomes same for the three data.

Finally, SIMS has been performed to investigate the validity of above theoretical finding. We prepared 3 samples: A dry-oxidized sample at 1300°C for 30 minutes, a dry-oxidized sample followed by the POCl$_3$ annealing at 1000°C for 10 minutes, and a dry-oxidized sample followed by the POCl$_3$ annealing at 1100°C for 3 minutes. In Fig.5, we show the evaluated profile of C concentration in SiO$_2$/SiC and PSG/SiC structures. When we compare profiles of the dry-oxidized and POCl$_3$-annealed samples, we observe that the C signals show the prominent tail for a few-nm deeper in PSG than in SiO$_2$, which is indicative of the creation of the –O$_3$PCO$_2$ in PSG near the interface. Moreover, if we compare samples with different POCl$_3$ annealing temperature (1000°C or 1100°C), C is more distributed deeply inside the PSG with annealing at higher temperature. This trend is semi-quantitatively consistent with the calculated Gibbs free energy (Figs. 4(a) and (b)) which shows that the remaining –O$_3$PCO$_2$ inside the PSG near the interface increases with higher POCl$_3$ annealing temperature.

In summary, we have performed the first-principle static and dynamic calculations, identified characteristic structure in P-doped SiO$_2$ as the –O$_3$PO configuration, and found that the configuration attracts CO units ejected from the SiO$_2$/SiC interface, thus operating as a carbon absorber during the SiC oxidation. This finding has provided the first microscopic reasoning for the reduction of the C-related interface states and the promotion of the oxidation rate through the phosphorous treatment. The SIMS measurements clearly corroborate this theoretical finding.

Computations were performed mainly at the Center for Computational Science, University of Tsukuba, and the Supercomputer Center at the Institute for Solid State Physics, The University of Tokyo. Y.M. acknowledges the support from JSPS Grant-in-Aid for Young Scientists (B) (Grant Number 16K18075). The experimental part (SIMS measurement) of the work was supported in part by the Super Cluster Program of the Japan Science and Technology Agency.

[25] The oxygen partial pressure $p$ and its chemical potential $\mu_O(O_2)$ are related in the equation, $\mu_O(O_2) = \mu_O^0(O_2) + \frac{kT}{2}\ln(\frac{p}{p^0})$, where $\mu_O^0(O_2)$ is the oxygen chemical potential at standard-state pressure and $k$ the Boltzmann's constant. By taking into account the solubility of 1-atm $O_2$ in $SiO_2$ at 1000°C (~ 2.1×10$^{16}$ cm$^{-3}$ [26]) and the mole fraction of $O_2$ (~ 0.16) during the $POCl_3$ annealing, one may estimate the partial pressure of $O_2$ in $SiO_2$ as ~ 61 Pa. We here assume that every P in $POCl_3$ completely becomes the form of $P_2O_5$. Then, $\mu_O^0(O_2)$ (the right end of the Fig.4(b)) shifts toward the negative direction by ~ 0.4 eV.

[26] K. Kajihara, H. Kamioka, T. Miura, L. Skuja, and H. Hosono, *J. Appl. Phys.* **98**, 013529 (2005).